# Experimental Verification of the Chemical Sensitivity of Two-Site Double Core-Hole States Formed by an X-ray FEL

P. Salén[1*], P. van der Meulen[1], H.T. Schmidt[1], R.D. Thomas[1], M. Larsson[1], R. Feifel[2], M.N. Piancastelli[2], L. Fang[3], B. Murphy[3], T. Osipov[3], N. Berrah[3], E. Kukk[4], K. Ueda[5], J.D. Bozek[6], C. Bostedt[6], S. Wada[6,7], R. Richter[8], V. Feyer[8], K.C. Prince[8,9]

[1]Stockholm University, Physics Department, 106 91 Stockholm, Sweden

[2]Uppsala University, Department of Physics and Astronomy, 751 20 Uppsala, Sweden

[3]Western Michigan University, Physics Department, Kalamazoo, MI 49008, USA

[4]University of Turku, Department of Physics and Astronomy, 20014 Turku, Finland

[5]IMRAM, Tohoku University, Sendai 980-8577, Japan

[6]SLAC, LCLS, Menlo Park, CA 94025, USA

[7]Hiroshima University, Department of Physical Science, Higashi-Hiroshima 739-8526, Japan

[8]Sincrotrone Trieste, 34149 Basovizza, Trieste, Italy

[9]IOM-CNR, 34149 Basovizza, Trieste, Italy

*email: peter.salen@fysik.su.se

**We have performed X-ray two-photon photoelectron spectroscopy (XTPPS) using the Linac Coherent Light Source (LCLS) X-ray free-electron laser (FEL) in order to study double core-hole (DCH) states of $CO_2$, $N_2O$ and $N_2$. The experiment verifies the theory behind the chemical sensitivity of two-site (ts) DCH states by comparing a set of small molecules with respect to the energy shift of the tsDCH state and by extracting the relevant parameters from this shift.**

The LCLS X-ray free-electron laser, at the SLAC National Accelerator Laboratory, produces ultra-short laser pulses with extremely high peak intensities in both the soft and hard X-ray domain [1,2]. These characteristics enable the exploration of hitherto virtually unmapped scientific territories, such as the non-linear interaction between matter and X-ray photons, and allows for a natural continuation of the already well established field of optical non-linear laser spectroscopy [3]. An intriguing example of such an X-ray-induced multiphoton process is the production of DCH states via the sequential absorption of two soft-X-ray photons on a time-scale on the order of the molecular Auger lifetime (~4-8 fs) [4]. The formation of molecular tsDCH states in particular shows great promise as a powerful tool for chemical analysis [5,6], and recently has attracted considerable attention [7-10]. The unique properties of the LCLS permit the search for these double core vacancies located at different atomic sites using XTPPS [11-14].

A compelling motivation for the study of tsDCH states is their ability to probe the local chemical environment more sensitively than either single core-hole (SCH) [15] or single-site (ss) DCH [16-18] states as predicted by Cederbaum *et al.* in their seminal paper from 25 years ago [5]. Their results were confirmed recently by Tashiro *et al.* [6] who calculated the single ionization potentials (IP) and double core-hole ionization potentials (DIP) for a series of small molecules. The increased sensitivity originates from the fact that the DIP of tsDCH states is directly coupled to induced changes in the valence charge distribution at the two different atomic sites [6]. Here we set out to verify these theoretical predictions by measuring the DIPs of the tsDCH states for a set of small molecules, viz. $N_2$, $N_2O$, $CO_2$ and CO [8].

The experiments reported here [19] were performed using the Atomic, Molecular and Optical (AMO) instrument [20,21] of the LCLS. The FEL generated light pulses with a FWHM (full width at half maximum) duration of ~10 fs, a photon energy of between 517 and 705 (± 15) eV, and a pulse energy of approximately 30 μJ on the target. A tightly focused laser beam provided the high intensity in the interaction region ($3\times10^{16}$ W/cm$^2$) that enabled sequential ionization of the molecules. Data taken with an unfocused beam ($1\times10^{14}$ W/cm$^2$) was subtracted from data taken with the focused beam in order to extract the non-linear contributions to the photoelectron signal, and more clearly observe the DCH states.

In the difference spectra a number of features can be discerned that are unambiguously related to the sequential absorption of two soft X-ray photons. First, at kinetic energies that are about 50-100 eV lower than the ordinary $1s^{-1}$ photoline (depending on the atom involved) a peak is observed that can be confidently assigned to the ssDCH state [6]. The tsDCH states are located much closer to the main photoline, typically shifted to lower energy by about 10-20 eV [6]. If the pulse duration exceeds the Auger lifetime, Auger decay can take place before the absorption of a second photon. This gives rise to the so-called Photoemission-Auger-Photoemission (PAP) peaks in the photoelectron spectra, whose location can be calculated from the energies of the relevant doubly and triply ionized states of the molecule [10,18,22-24]. Generally, PAP peaks appear at kinetic energies of about 20-40 eV lower than the main photoline.

The relative intensities of the ssDCH, tsDCH and PAP peaks, as well as that of the main photoline, can be simulated on the basis of a straightforward kinetic model which has been shown to produce reliable results [25]. For our experimental conditions and for the molecules studied here, the various ssDCH and tsDCH peaks for a particular molecule are expected to have very similar integrated intensities, typically within a factor of 2 [19]. This means that the presence of the usually easily identifiable ssDCH peak implies the existence of a roughly equally intense tsDCH peak. This is of great help in analyzing the spectra because the latter generally lies in a more congested region of the photoelectron spectrum. The relative intensity of the PAP peaks is calculated to be comparable to that of the DCH lines [19].

The most prominent DCH structures were observed in $N_2$ and we begin the discussion with $N_2$ and $N_2O$, before we continue with our observations of DCHs in $CO_2$ and CO [8].

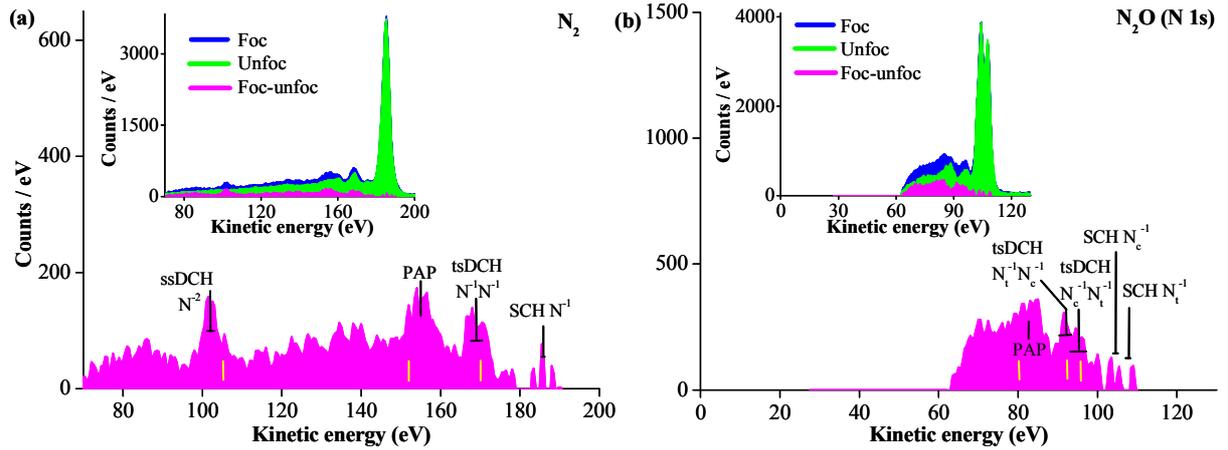

FIG. 1 (color). Photoelectron spectra of $N_2$ (a) and $N_2O$ (b) with a photon energy of 596 eV and 517 eV, respectively. Blue curve: focused X-rays; green curve unfocused; spectra are normalized at the SCH peak. Magenta curve: difference spectra. Black vertical lines: experimental energies of the SCH, DCH and PAP features; horizontal lines: estimated errors; yellow vertical lines: theoretical energies [6].

Figure 1(a) shows the photoelectron spectra of $N_2$ at kinetic energies below the N SCH peak at 186 eV, where the DCH features are expected to lie. The photon energy was 596 eV and the electron time-of-flight (eTOF) spectrometer was oriented at the magic angle (54.7° with respect to the polarization of the laser beam). The blue and green spectra correspond to data taken with the focused and unfocused beams, respectively, and are normalized to the N SCH peak. The magenta spectrum plots the difference between these data to highlight non-linear contributions to the signal, i.e. the DCHs. The experimentally determined positions of the non-linear peaks are indicated in Fig. 1(a) by the black vertical lines (the estimated error is represented by the horizontal line) together with the theoretically predicted positions [6] (yellow vertical lines). The energy resolution is ~4.6 eV given by the FWHM of the N SCH peak.

Data obtained with the unfocused beam are free from non-linear contributions and comparable to data taken using synchrotron light [19], consisting of the N SCH peak and its shake-up and shake-off satellites. The largest satellite intensity is expected at kinetic energies of about 155-175 eV [26,27] and such features are observed in both the focused and unfocused spectra (Fig. 1(a)). At lower electron kinetic energies one expects to find shake-off and smaller satellites.

The difference spectrum in Fig. 1(a) clearly reveals three distinct features located at kinetic energies of ~102, ~155, and ~169 eV. These correspond to the ssDCH, PAP and tsDCH states, respectively. A continuous background also appears in this difference spectrum, as well as in the spectra of the other measured molecules. Its origin is not entirely clear but the background may partly arise from atomic ions produced by rapid fragmentation [8], satellites to the PAP and DCH peaks, or may be related to secondary electrons that are not completely eliminated in the subtraction procedure due to the different source volumes in the focused and unfocused measurements.

Figure 1(b) shows the corresponding photoelectron spectra for $N_2O$. The photon energy in these measurements was 517 eV, i.e. below the ionization limit of O 1s electrons in order to generate signals solely from N related DCHs. This facilitates easier comparison between the $N^{-1}N^{-1}$ tsDCH in $N_2$ and $N_2O$. (For the photoelectron spectra of $N_2O$ obtained using a photon energy of 597 eV, i.e. above the O 1s ionization limit, see supplemental material [19]). The kinetic-energy region of interest is smaller for $N_2O$ than $N_2$, due to the reduced photon energy, but for ease of comparison the same kinetic energy-scale as in Fig. 1(a) is used. The reduced photon energy also results in an increased energy resolution, ~4 eV, compared with that observed for $N_2$.

The asymmetry of $N_2O$, with one central ($N_c$) and one terminal ($N_t$) N atom, manifests itself in a splitting of the N SCH peak of about 4.0 eV. The difference spectrum shows similar structures to those observed in $N_2$: a tsDCH peak followed by a PAP peak. The position of the experimentally observed tsDCH peak is in good agreement with the calculated value [6]. This peak consists of contributions from the $N_c^{-1}N_t^{-1}$ ($N_c$ followed by $N_t$ core-ionization) and $N_t^{-1}N_c^{-1}$ ($N_t$ followed by $N_c$ core-ionization) state, which seem to be just resolved. The marked experimental positions are also reasonable given the expected separation of 4 eV between the two peaks.

TABLE 1. Experimental and theoretical ionization potentials as well as $\Delta E_1$, $\Delta E_2$ and the generalized interatomic relaxation energy, IRC [6], (see text) for states of $N_2$, $N_2O$, $CO_2$ and CO. IP($S^{-1}$): SCH ionization potential, DIP($S^{-2}$): ssDCH double ionization potential, DIP($S_i^{-1}$, $S_j^{-1}$): tsDCH double ionization potential. Experimental DIPs are calculated as the sum of the experimental IP($S^{-1}$), calibrated to known values, and the relevant IP of the ion, determined in our experiments. The experimental value of the IP for CO (O 1s) = 542.5 eV was taken from reference 28. Errors are calculated as the root mean square of the estimated errors for the positions of the spectral peaks.

| Molecule | IP($S^{-1}$) (eV) | DIP($S^{-2}$) (eV) | DIP($S_i^{-1}$,$S_j^{-1}$) (eV) | $\Delta E_1$($S^{-2}$) (eV) | $\Delta E_2$($S_i^{-1}$,$S_j^{-1}$) (eV) | IRC (eV) |
|---|---|---|---|---|---|---|
| $N_2$ | ($N^{-1}$) | ($N^{-2}$) | ($N^{-1}N^{-1}$) | ($N^{-2}$) | ($N^{-1}N^{-1}$) | ($N^{-1}N^{-1}$) |
| Exp. | 409.9±0.3 [29] | 903.2±1.1 | 836.2±1.6 | 83.4±1.1 | 16.4±1.6 | -3.29±1.6 |
| *Theory [6]* | *411.0* | *901.2* | *836.4* | *79.2* | *14.3* | *-0.65* |
| $N_2O$ ($N_t$) | ($N_t^{-1}$) | ($N_t^{-2}$) | ($N_c^{-1}N_t^{-1}$) | ($N_t^{-2}$) | ($N_c^{-1}N_t^{-1}$) | ($N_c^{-1}N_t^{-1}$) |
| Exp. | 409.0±0.5 | | 834.2±2.1 | | 12.7±2.1 | 0.09±2.1 |
| *Theory [6]* | *408.6* | *893.9* | *833.2* | *76.7* | *12.1* | *1.11* |
| $N_2O$ ($N_c$) | ($N_c^{-1}$) | ($N_c^{-2}$) | ($N_t^{-1}N_c^{-1}$) | ($N_c^{-2}$) | ($N_t^{-1}N_c^{-1}$) | ($N_t^{-1}N_c^{-1}$) |
| Exp. | 412.5±0.5 [29] | | 834.2±1.6 | | 12.7±1.6 | 0.09±1.6 |
| *Theory [6]* | *412.5* | *902.3* | *833.2* | *77.3* | *12.1* | *1.11* |
| $CO_2$ (O 1s) | ($O^{-1}$) | ($O^{-2}$) | | ($O^{-2}$) | ($O^{-1}O^{-1}$+$C^{-1}O^{-1}$) | |
| Exp. | 540.6±0.5 [30] | 1173.2±1.6 | | 92.0±1.5 | 12.8±1.6 | |
| *Theory [6]* | *542.9* | *1171.9* | | *86.2* | *9.1* | |
| $CO_2$ (C 1s) | ($C^{-1}$) | ($C^{-2}$) | ($O^{-1}C^{-1}$) | ($C^{-2}$) | ($O^{-1}C^{-1}$) | ($O^{-1}C^{-1}$) |
| Exp. | 296.8±0.5 [30] | | 848.6±1.6 | | 11.2±1.6 | 1.21±1.6 |
| *Theory [6]* | *297.6* | *664.6* | *851.2* | *69.3* | *10.6* | *1.79* |
| CO (C 1s) | ($C^{-1}$) | ($C^{-2}$) | ($O^{-1}C^{-1}$) | ($C^{-2}$) | ($O^{-1}C^{-1}$) | ($O^{-1}C^{-1}$) |
| Exp. [8] | 296.5±0.5 | 667.9±3.6 | 855.3±1.2 | 74.9±4.0 | 16.3±1.2 | -3.53±1.2 |
| *Theory [6]* | *296.4* | *664.4* | *855.2* | *71.7* | *15.9* | *-2.8* |

Table 1 lists the single (IP) and double ionization potentials (DIP) for the SCH and DCH states, respectively. The DIPs are derived from the energy shifts of the ssDCH ($\Delta E_1$) and tsDCH ($\Delta E_2$) states, where $\Delta E_1$ = DIP($S^{-2}$) - 2·IP($S^{-1}$) and $\Delta E_2$ = DIP($S_i^{-1}$,$S_j^{-1}$) - IP($S_i^{-1}$) - IP($S_j^{-1}$), which are directly determined from the spectra as the shift with respect to the SCH line. These values are also listed together with the DIPs and they all agree reasonably well with their corresponding theoretical values. The chemical sensitivity of the tsDCH states is reflected in the value of $\Delta E_2$, and we determine that $\Delta E_2$ is lower by 3.7 ± 2.7 eV for $N_c^{-1}N_t^{-1}$ in $N_2O$ compared with the $N^{-1}N^{-1}$ state in $N_2$, in tolerable agreement with the calculated difference (2.2 eV) [6]. This lowering is due to the influence of the extra O atom.

The predicted high sensitivity of the tsDCHs to the chemical environment is related to the flow of electron density in the creation of the tsDCH states [6]. This change of electron density may be quantified by the generalized interatomic relaxation energy (IRC) [6], which is related to $\Delta E_2$ according to $\Delta E_2 = 1/r - IRC$, where r is the distance between the two core holes. The decrease in $\Delta E_2$ of $N_2O$ and $CO_2$ with respect to $N_2$ and CO due to an extra O atom attached to each molecule is approximately 0.4 eV [31]. Values of IRC are given in Table 1. Theory predicts a modest difference between the IRC of the $N^{-1}N^{-1}$ tsDCH in $N_2$ and $N_2O$. This is manifested experimentally by the relatively small difference of $\Delta E_2$ (3.7 eV) between the two states, which is only slightly larger than the difference in the IPs of N in $N_2$ and $N_2O$, respectively. The effect is, however, more pronounced in the case of CO and $CO_2$, as we will show later.

For diatomic molecules, creation of the core hole decreases the electron density on the other atom and thus the relaxation is suppressed for the core-hole creation of the second atom, resulting in a negative IRC. For triatomic molecules, in which one of the core holes is located at the center atom, the extra atom plays the role of an electron donor to the other two atoms with core holes and enhances the relaxation of the double core hole at the other two sites, resulting in a positive IRC. We find that the relaxation is suppressed (IRC < 0) for $N_2$ and enhanced (IRC > 0) for $N_2O$, as predicted [6].

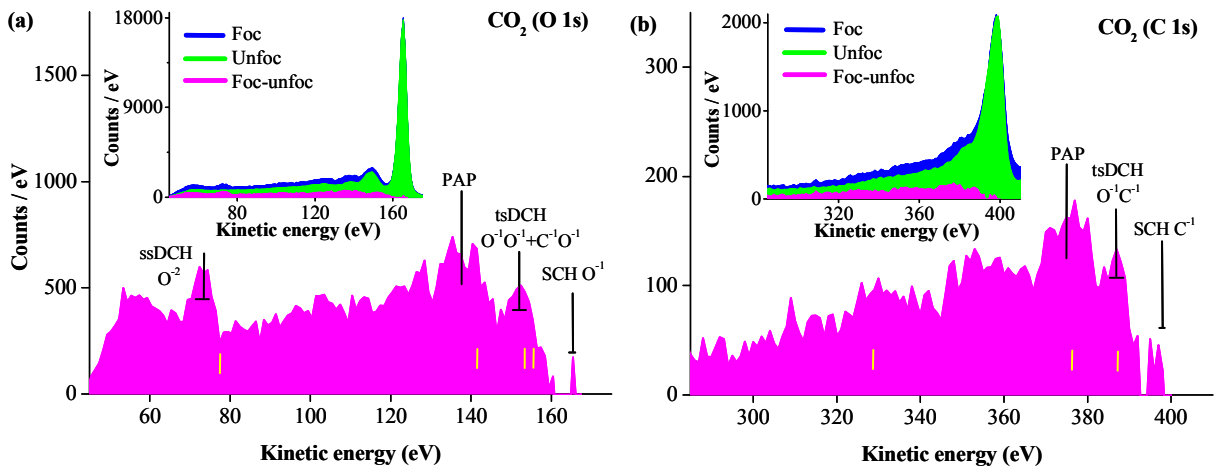

FIG. 2 (color). Photoelectron spectra of $CO_2$ near the O SCH line (a) and C SCH line (b), photon energy 705 eV. Blue curve: focused X-rays; green curve: unfocused X-rays; spectra are normalized at the SCH peak. Magenta curve: difference spectrum. Black vertical lines: experimental energies of the SCH, DCH and PAP features; horizontal lines: estimated errors; yellow vertical lines: theoretical energies [6].

Comparison between data obtained for CO and $CO_2$ allows the influence of the extra O atom in $CO_2$ on the $O^{-1}C^{-1}$ tsDCH state to be investigated. The photoelectron spectra of CO, in which the $O^{-1}C^{-1}$ tsDCH was positively identified, was recently reported by Berrah *et al.* [8], and we here present the data obtained for $CO_2$.

The measurements on $CO_2$ were recorded with an eTOF oriented parallel to the polarization of the laser. As in Fig. 1, Fig. 2(a) shows the photoelectron spectra of $CO_2$ for kinetic energies below the O SCH peak at 165 eV and were obtained using a photon energy of 705 eV. The assignments of the peaks are marked in the figure. For the peak at ~152 eV, theory predicts contributions from two close lying tsDCH states separated by ~2 eV [6]. One arises when a 1s electron is ejected from each O atom ($O^{-1}O^{-1}$) and the other when the first electron originates from the C atom and the second from one of the O atoms ($C^{-1}O^{-1}$). However, the energy resolution here is ~5 eV, and we do not resolve these contributions.

The spectra shown in Fig. 2(b) were taken in the same measurement as Fig. 2(a) but now we highlight the kinetic energy region near the C SCH peak. Here one expects to find only one tsDCH peak, arising from the $O^{-1}C^{-1}$ state. This peak appears at the calculated position in the difference spectrum (see Fig. 2(b)). The PAP peak is also observed, but the carbon $(1s)^{-2}$ ssDCH peak cannot be positively identified. This is in full agreement with our simulations which indicate a much reduced intensity for this particular feature [19].

The DCH features for $N_2O$ and $CO_2$ are less pronounced than in the case of $N_2$ due to the operating conditions of the LCLS at the time these measurements were made. However, our assignments are supported strongly not only by the calculated peak positions [6,8,18,22-24] but also by our simulations of the expected relative intensities of the ssDCH, PAP, tsDCH and SCH peaks [19]. In addition, the experimental results for $N_2$ clearly establish the principle, and the pattern with ssDCH, PAP and tsDCH peaks is repeated for all molecules.

Table 1 lists the IPs and DIPs for the SCH and DCH states together with $\Delta E_1$ and $\Delta E_2$ for $CO_2$ and CO. Table 1 also lists the results from both the O 1s and C 1s peaks of $CO_2$, although the observed tsDCH peak adjacent to the O 1s line cannot be resolved into its two components, and is not used for comparison with CO. The tsDCH peak close to the C 1s line consists of only one contribution, that from $O^{-1}C^{-1}$, and is suitable for extracting $\Delta E_2$. Comparing $\Delta E_2$ for the $O^{-1}C^{-1}$ tsDCH state of CO and $CO_2$ allows us to evaluate experimentally the environmental effect of the extra O atom in $CO_2$. Here, $\Delta E_2$ for the $O^{-1}C^{-1}$ tsDCH state of $CO_2$ is 5.1 ± 2.3 eV lower than that of CO, in good agreement with the predicted value of 5.3 eV [6]. It is noted that the IP difference between $C^{-1}$ SCH states in CO and $CO_2$ is only 0.3 eV, while the DIP difference between the $O^{-1}C^{-1}$ tsDCH states is 6.7 eV. The higher sensitivity of tsDCH states to the chemical environment is evident.

If we focus on the IRC values given in table 1 we find that the large difference in IRC between the $O^{-1}C^{-1}$ state of CO and $CO_2$, compared with the case of $N_2$ and $N_2O$, which results in the large shift of $\Delta E_2$ in the former case, fits nicely with theory [6]. In addition we find also for these two molecules that the relaxation is suppressed (IRC < 0) for the diatomic CO and enhanced (IRC > 0) for the triatomic $CO_2$, as predicted [6].

In conclusion, we have presented evidence for the formation of tsDCH states in the molecules $N_2$, $N_2O$ and $CO_2$ by employing photoelectron spectroscopy using a FEL X-ray light source. Our experimental results for the DIP, $\Delta E_2$ and IRC reproduce the trends predicted by the theory for tsDCH states [6], and thus support its main implication that the tsDCH states are extra sensitive to the chemical environment. In particular the IRC, a characteristic parameter of the tsDCH states, was found to behave according to theory. The sensitivity of the tsDCH state to the chemical environment was exemplified by the large spectral shift of the $O^{-1}C^{-1}$ tsDCH state in $CO_2$ compared with CO, which is induced by the extra O atom.

**Acknowledgments**

This work was funded in part by DOE, Office of Science, Basic Energy Science, Chemical, Geosciences, and Biological Divisions. Funding from MEXT, JST, JSPS, Japan, MIUR Italy (grants FIRB-RBAP045JF2, FIRB-RBAP06AWK3) and the Swedish Research Council (VR) is gratefully acknowledged. We thank M. Tashiro, M. Ehara and P. Juranic for their participation and all of the LCLS support staff, in particular J. C. Castagna and M. L. Swiggers.